\documentclass[12pt]{article}
\usepackage{amssymb}
\usepackage{epsfig}
\usepackage{amssymb,amsmath}
\usepackage{slashed}
\usepackage{multirow}

\newcommand{\bea}{\begin{eqnarray}}
\newcommand{\eea}{\end{eqnarray}}

 \makeatletter
\def\alt{\mathrel{\mathpalette\gl@align<}}
\def\agt{\mathrel{\mathpalette\gl@align>}}
\def\gl@align#1#2{\lower.6ex\vbox{\baselineskip\z@skip\lineskip\z@
\ialign{$\m@th#1\hfil##\hfil$\crcr#2\crcr\sim\crcr}}} \makeatother

\begin{document}
%
\vspace*{1.0cm}

\begin{center}
\baselineskip 20pt {\Large\bf Amelioration of Little Hierarchy Problem  in $SU(4)_c
\times SU(2)_L \times SU(2)_R$} \vspace{1cm}

{\large  Ilia Gogoladze \footnote{ On  leave of absence from:
Andronikashvili Institute of Physics,  Tbilisi, Georgia.
\\ \hspace*{0.5cm} }, Mansoor Ur Rehman and Qaisar
Shafi} \vspace{.5cm}

{\baselineskip 20pt \it Bartol Research Institute, Department of Physics and Astronomy, \\
University of Delaware, Newark, DE 19716, USA \\

 }
\vspace{.5cm}

\end{center}

\begin{abstract}

The little hierarchy problem encountered in the constrained minimal
supersymmetric model (CMSSM) can be ameliorated in
supersymmetric models based on  the gauge symmetry $G_{422} \equiv SU(4)_c \times
SU(2)_L \times SU(2)_R$. The standard assumption in CMSSM (and in SU(5) and SO(10))
of universal gaugino masses can be relaxed in $G_{422}$,
and this leads to a significant improvement in the degree of fine tuning required to implement
radiative electroweak breaking in the presence of a characteristic
supersymmetry breaking scale of around a TeV. Examples of Higgs and sparticle
mass spectra realized with 10\% fine tuning are presented.
\end{abstract}
\date{}

\newpage

\section{Introduction}

The constrained minimal supersymmetric standard model (CMSSM) based
on supergravity \cite{Chamseddine:1982jx} with R-parity
conservation is a well motivated extension of physics beyond the
Standard Model (SM). Among other things it predicts that 
the lightest CP-even Higgs boson mass, after including 
radiative corrections, is $m_h \lesssim 125$ GeV
\cite{at, Carena:1995wu}. This is to be compared with the LEP2 lower bound 
$m_{h}\geq $ $114.4$ GeV \cite{LEP2}. Values of $m_{h}$  around $125$ GeV 
or so require that the soft supersymmetry (susy) breaking parameters 
are of order a TeV scale.
Such values, in turn, lead to the so-called little
hierarchy problem \cite{b5} because in implementing radiative
electroweak symmetry breaking, TeV scale quantities must conspire to
yield the electroweak mass scale $M_{Z}$.

A variety of scenarios have been proposed \cite{casas,
Babu:2008ge} to solve this so-called little hierarchy
problem. Many of them extend the gauge and/or matter sector of the CMSSM in
order to increase $m_h$ while keeping the SUSY
particle mass  spectrum as light as possible. It has been shown in
\cite{Abe:2007kf} that non-universal gaugino masses at the GUT scale 
$M_G$ can help resolve the little hierarchy problem. To do this 
the authors have studied a variety of gaugino mass ratios 
obtained from some underlying theories.
Following up on this, in this paper we investigate $SU(5)$ and 
$SO(10)$ GUTs inwhich non-universal gaugino
masses are realized via dimension five operators. We focus, in particular, 
on the $SU(4)_c \times SU(2)_L \times SU(2)_R$ ($G_{422}$) model \cite{Pati:1974yy} which 
provides a natural setup for non-universal gaugino masses. In the $G_{422}$ model
the little hierarchy problem  can be largely resolved if $SU(2)_L$ and $SU(3)_c$
gaugino masses satisfy the asymptotic relation $M_2/M_3\approx 4$.

The plan of the paper is as follows. In  Section 2 we review the
little hierarchy problem in the CMSSM with universal gaugino masses. In
Section 3 we compare the fine tuning in $SO(10)$ and $SU(5)$
GUT models with non-universal gaugino masses induced by suitable dimension five 
operators. In Section 4 we discuss the $G_{422}$ model where we present the solution 
of the little hierarchy problem as well as its implications for the sparticle 
and Higgs spectrum. Our conclusions are summarized in Section 5.

\section{Little Hierarchy Problem in MSSM}
%
%

\begin{figure}[t]
\centering \includegraphics[angle=0, width=13cm]{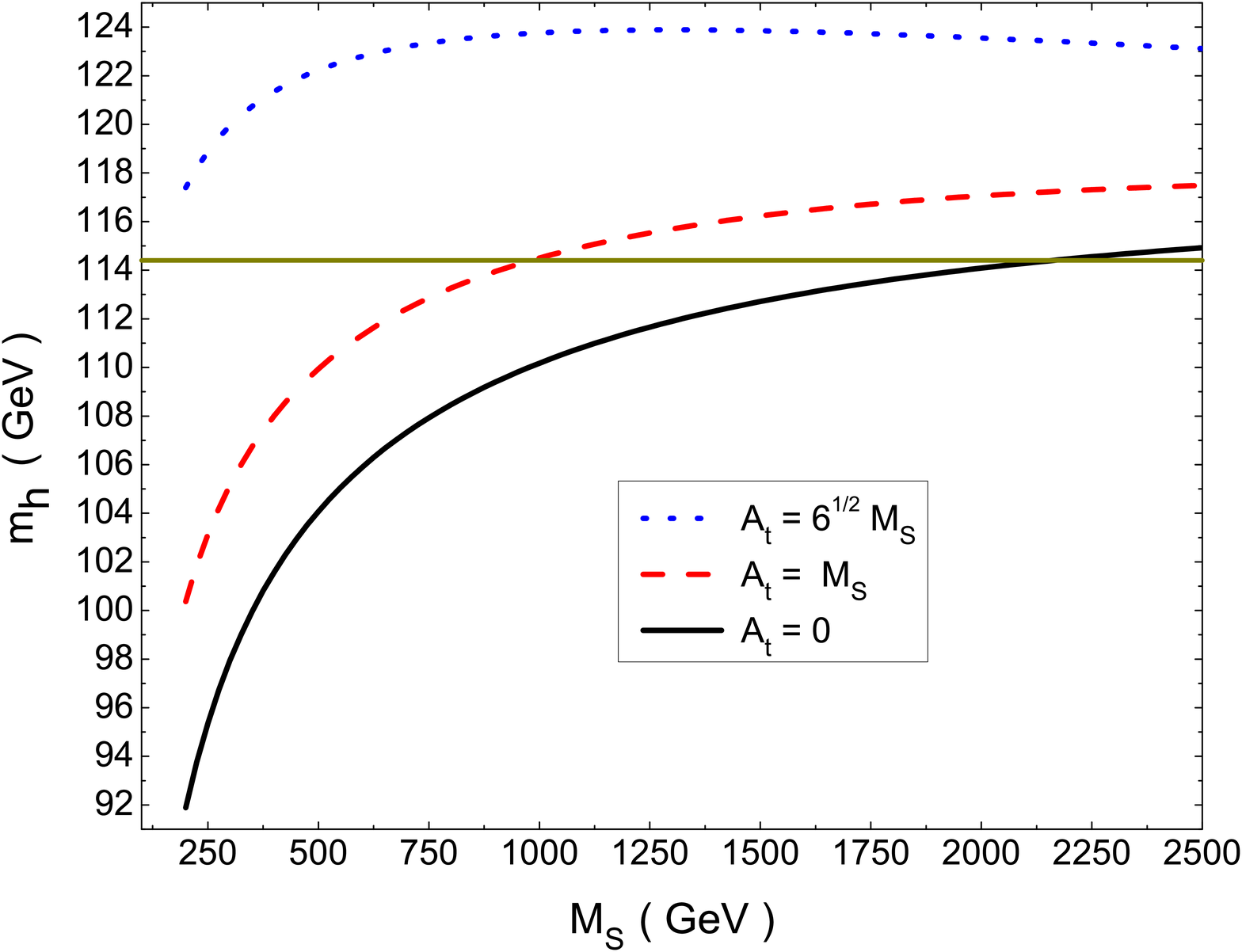}
\vspace{-1cm} \caption{Higgs mass $m_h$ vs $M_{S}$ for different maximal and minimal values of
$A_{t}/M_{S}$, with  tan$\beta$ = 10 and $M_t=172.6$ GeV. Solid,
dashed and dotted  lines correspond to $A_t=0, M_S$ and $\sqrt6 \,M_S$
respectively. The horizontal line denotes the LEP2 bound $m_h \geq 114.4$
GeV. } \label{MSvsMH}
\end{figure}

At tree level the lightest CP even Higgs boson mass $m_h$ in MSSM is bounded from above
by the mass of the $Z$ boson
\begin{equation}
m_h \leq M_Z.
\end{equation}
Thus, significant radiative corrections are needed in order to push lightest CP even Higgs boson mass
above the LEP2 limit $m_h \geq 114.4$ GeV. One finds \cite{at, Carena:1995wu}
\begin{eqnarray}
m_{h}^{2} & \simeq & M_{Z}^{2}\cos ^{2}2\beta \left( 1-\frac{3%
}{8\pi ^{2}}\frac{m_{t}^{2}}{v^{2}}t\right)  \nonumber \\
&&+\frac{3}{4\pi ^{2}}\frac{m_{t}^{4}}{v^{2}}\left[ t+\frac{1}{2}X_{t}+\frac{%
1}{\left( 4\pi \right) ^{2}}\left(
\frac{3}{2}\frac{m_{t}^{2}}{v^{2}}-32\pi \alpha _{s}\right) \left(
X_{t}t+t^{2}\right) \right],  \label{e1}
\end{eqnarray}%
where
\begin{eqnarray}
t &=&\log \left( \frac{M_{S}^{2}}{M_{t}^{2}}\right), X_{t} =
\frac{2\widetilde{A}_{t}^{2}}{M_{S}^{2}}\left( 1-\frac{\widetilde{A}%
_{t}^{2}}{12M_{S}^{2}}\right),
\end{eqnarray}%
$\widetilde{A}_{t}=A_{t}-\mu \cot \beta $, where $A_{t}$
denotes the stop left and stop right soft mixing parameter, $\mu$ is
the MSSM Higgs bilinear mixing term, and
$M_S^2=\sqrt{m^2_{Q_t}m^2_{U_t}}$ is the geometric mean of left and right
stop masses squared. As seen from Figure \ref{MSvsMH}, if the
Higgs mass $m_h$ turns out to be significantly larger than 114.4 GeV, one would require
$\vert A_t \vert \gtrsim M_S$, or relatively heavy stop quarks 
($M_S \gtrsim $ 1 TeV), or a suitable combination of the two.

The $Z$ boson mass is obtained from the
minimization of the Higgs scalar potential, and for $\tan\beta
\gtrsim 10$, we have the standard approximate relation
\begin{equation}
\frac{1}{2}M_{Z}^{2}\approx -\mu (M_{Z})
^{2}+\left(\frac{m_{H_{d}}^{2}(M_{Z}) - m_{H_{u}}^{2} (M_{Z}) \text{
tan}^{2}\beta}{\text{ tan}^{2}\beta -1} \right)\simeq -\mu (M_{Z})
^{2}-m_{H_{u}}^{2} (M_{Z}). \label{e4}
\end{equation}
In order to see the explicit dependence of $m_{H_{u}}^{2} (M_{Z})$
on $A_t$ and $M_S$ we employ a semi-analytic expression for 
$m_{H_{u}}^{2} (M_{Z})$ using one loop renormalization group 
equations (RGEs) \cite{RGE} in the $\overline{DR}$ regularization 
scheme for the soft supersymmetry breaking terms and
the MSSM gauge couplings. We take the GUT scale $M_G$ $\simeq 2.0
\times 10^{16}$ GeV, and  $\alpha_2 = \alpha_1 = \alpha_G =
1/24.32$. We do not enforce exact unification $\alpha_3 = \alpha_2 =
\alpha_1$ at $M_G$, since a few percent deviation from the
unification condition can be expected due to unknown GUT scale threshold
corrections \cite{Hisano:1992jj}. Including only the dominant terms,
the semi-analytic expression for $m_{H_{u}}^{2} (M_{Z})$ in CMSSM 
can be written as
\begin{equation}
- m_{H_{u}}^{2} (M_{Z}) \simeq   1.05\,A_t^2 + 1.39\,A_t\,M_S + 0.87\,M_S^2 = 
\left( 1.05\,\frac{A_t^2}{M_S^2} +1.39\,\frac{A_t}{M_S} + 0.87 \right)\,M_S^2 .
 \label{mhu}
\end{equation}%
For $m_h \gtrsim 114.4$ GeV, which is realized with either $M_S \gtrsim $ 1 TeV and/or
$\vert A_t \vert \gtrsim M_S$, $\vert m_{H_{u}} \vert$ is much larger than 
$M_Z$ which, in turn, implies that $\mu$ should be fine tuned in order to achieve
the correct EW symmetry breaking in Eq.(\ref{e4}). 
This fine tuning is referred to as the little hierarchy problem.
Following the analysis in \cite{{Barbieri:1987fn},Abe:2007kf}, we 
consider a quantitative measure of fine tuning 
\begin{eqnarray}
\Delta_X &=& \frac{1}{2}\frac{X}{M_Z^2} \frac{\partial
M_Z^2}{\partial X}, \qquad (X=\mu_0,\, M_1,\, M_2,\, M_3,\, A_{t_0}, \, m_0), \label{ftdef}
\end{eqnarray}
where $M_{1,2,3}$, $A_{t_0}$ and $m_0$ are respectively the 
gaugino masses, trilinear coupling and the universal soft scalar mass at 
$M_G$. In the CMSSM the dominant contribution is given as
\begin{equation}
M_{Z}^{2} \simeq -2.04\,\mu_0^{2}+5.44 \,m_{1/2}^{2}+0.183 \, m_{0}^{2}+0.2\,
A_{t_{0}}^{2}-0.87 \,m_{1/2}A_{t_{0}},\label{MZ1}
\end{equation}
with $M_1 = M_2 = M_3 = m_{1/2}$. The definition of $\Delta_X$
is consistent with the normalization, 
\begin{eqnarray}
\sum_X \Delta_X &=& 1. \label{eq:ewconstraint}
\end{eqnarray}
\begin{figure}[t]
\centering \includegraphics[angle=0, width=13cm]{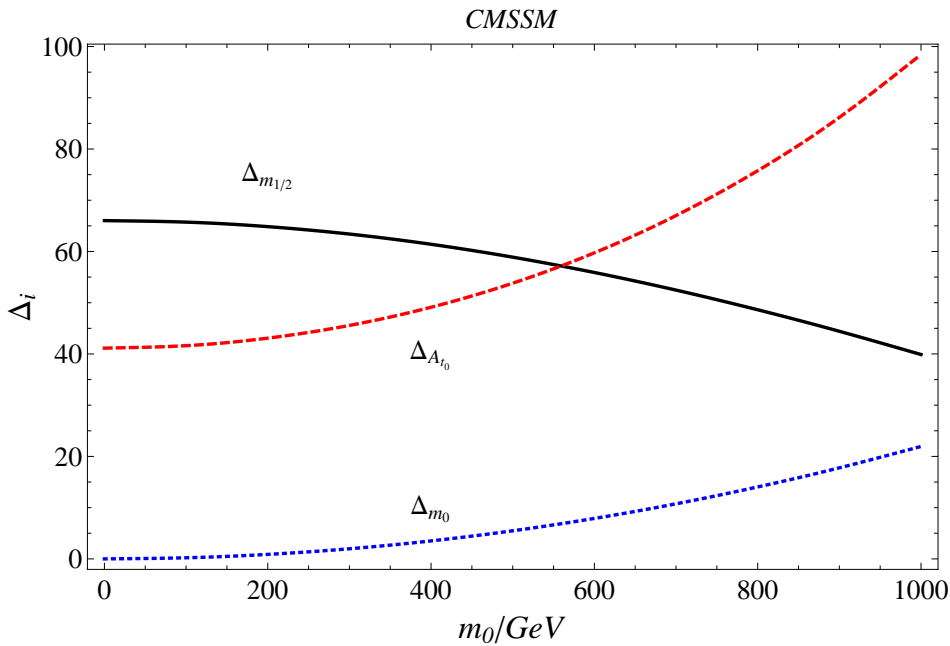}
\caption{$\Delta_i$ vs $m_0$ for $M_{S}=500$ GeV, $m_h = 119$ GeV,  tan$\beta$ = 10 
and $M_t=172.6$ GeV.} \label{FP}
\end{figure}
The quantity $\Delta_X$ measures the sensitvity of the $Z$-boson mass 
to the parameter $X$ (at $M_G$), such that $\left( \frac{100}{max(\vert \Delta_{X} \vert )}\right) \%$
is the degree of fine tuning. Normally, 10\% or somewhat large value of fine tuning is
regarded as the acceptable, and therefore we require max$(\vert \Delta_{X} \vert ) \lesssim 10$. 
For instance, with $m_h=119$ GeV, $\tan\beta=10$, $M_S=500$ GeV and $m_0 = 0$, as shown 
in the Figure \ref{FP}, the degree of fine tuning in CMSSM $\sim 1.5\%$, and
it gets worse with an increase in $M_S$. 
It is also interesting to view the dependence of the sensitivity parameters on $m_0$,
as shown in Figure \ref{FP}. For lower values of $m_0$, max$(\vert \Delta_{X} \vert )=\Delta_{m_{1/2}}$, 
while for higher values of $m_0$, max$(\vert \Delta_{X} \vert )= \vert\Delta_{A_{t_0}} \vert$. 
We see from Eq.(\ref{MZ1}) and Figure \ref{FP} that both $M_Z^2$ and max$(\vert \Delta_{X} \vert )$
have a weak dependence on the universal scalar mass $m_0$, provided $m_0 < $ TeV. 
Unless explicitly stated otherwise, we set $m_0 = 0$ in our numerical analysis. 
As suggested in ref. \cite{Abe:2007kf}, we can reduce the degree of fine 
tuning to an acceptable level by
employing suitable non-universal asymptotic gaugino masses. This non-universality
can be generated in GUTs from dimension five operators, but
it can be present in the $G_{422}$ model by a straightforward assumption.
In the next two sections, after discussing the origin of non-universality, we 
study the status of little hierarchy problem in $SU(5)$, $SO(10)$ and 
especially in the $G_{422}$ model.

\section{Little Hierarchy Problem in  $SU(5)$ and $SO(10)$}

A chiral superfield whose scalar component spontaneously breaks a GUT symmetry
(such as $SU(5)$ or $SO(10)$) may also possess a
 non-vanishing  $F$-component, in which case supersymmetry can also be broken \cite{Ovrut:1984qj}.
 In this section we briefly summarize  how this may lead to non-universal
 gaugino masses at $M_G$. In turn, this can help ameliorate the little hierarchy problem.
Consider the following dimension five operator for generating the soft gaugino
mass terms:
\begin{equation}
\int d^2 \theta  W^aW^b \left( \frac{\Phi_{ab}}{m_P} \right)+h.c. \supset \frac{\langle F_{\Phi}\rangle_{ab}}{m_P}
 \lambda^{a} \lambda^{b} + h.c. \,.    \label{ng}
\end{equation}
Here $W^a$ is the supersymmetric gauge field strength, $F_{\Phi}$  is the auxiliary field component of the chiral superfield $\Phi$, and $\lambda_{a}$ denotes the gaugino field. The field  $F_{\Phi}$, whose vacuum expectation value (VEV) breaks supersymmetry, is a singlet under the MSSM gauge group.
In $SU(5)$, for example, $\Phi$ can belong to
one of the following representations:
\begin{eqnarray}
(24\times 24)_{sym}=1+24+75+200, \label{non1}
\end{eqnarray}
while for $SO(10)$
 \begin{eqnarray}
(45\times 45)_{sym}=1+54+210+770.
\label{non2}
\end{eqnarray}
In this paper we will employ the widely used $24$ and 
$75$ dimensional representations  of $SU(5)$, and the $54$
dimensional representation of $SO(10)$.

In Table \ref{tab:1} we display the ratios among gaugino masses at
scales $M_{G}$ and $M_{Z}$. In $54$-plet the $SO(10)$ breaking
proceeds via $G_{422}$ \cite{Kibble:1982ae}. As seen from Table
\ref{tab:1}, the 24-plet of $SU(5)$ and the 54-plet of
$SO(10)$ yield identical  gaugino mass ratios at $M_G$ 
\cite{Martin:2009ad}, and therefore the results, as far as the 
little hierarchy problem is concerned, are the same for these two cases.
\begin{table}[t]
\begin{center}
\begin{tabular}{|c|c|c|}
\hline
SU(5) Representation & $M_{1}:M_{2}:M_{3}$ at $M_{G}$ & $M_{1}:M_{2}:M_{3}$ at $M_{Z}$  \\
\hline
{\bf 1} & 1:1:1 & 1:2:7 \\
\hline
{\bf 24} & (-1):(-3):2 & (-1):(-6):13.8\\
\hline
{\bf 75} & (-5):3:1 & (-1):1.2:1.4\\
\hline \hline
SO(10) Representation & $M_{1} :M_{2}:M_{3}$ at $M_{G}$ & $M_{1}:M_{2}:M_{3}$ at $M_{Z}$  \\
\hline
{\bf 1} & 1:1:1 &  1:2:7\\
\hline
{\bf 54}: {$ SU(4)_c \times SU(2)_L \times SU(2)_R$} & (-1):(-3):2
 & (-1):(-6):13.8\\
\hline
\end{tabular}
\end{center}
\caption{Gaugino mass ratios
 for $SU(5)$ and $SO(10)$ \cite{Martin:2009ad} at GUT and
electroweak scales.} \label{tab:1}
\end{table}

Before discussing the effects of specific choices of non-universal gaugino masses
on the little hierarchy problem we present a more general analysis for arbitrary gaugino
masses at $M_G$. For this purpose we perform a semi-analytic calculation for the MSSM
sparticle spectra with the following boundary
conditions
\begin{equation}
\{\alpha _{G},\,\, M_{G},\,\, y_{t}(M_{G})\} \approx  \{{1}/{24.32},\,\,
2.0\times 10^{16},\,\, 0.512\}.
\end{equation}
By integrating the one loop RGEs \cite{RGE},
we express the MSSM sparticle masses at scale $M_{Z}$ in terms
of the GUT scale fundamental parameters ($m_0, M_{1,2,3},
A_{t_{0}}$) and the Higgs bi-linear mixing term $\mu$. For example, the gaugino
masses at $M_{Z}$ are given by
\begin{equation}
\{M_{1}\left( M_{Z}\right) ,M_{2}\left( M_{Z}\right) ,M_{3}\left(
M_{Z}\right) \} \approx \{0.412M_{1},0.822M_{2},2.844M_{3}\}.
\label{h0}
\end{equation}%

Similarly,
\begin{eqnarray}
-m_{H_{u}}^{2}\left( M_{Z}\right)
& \simeq &2.67 M_{3}^{2}-0.2 M_{2}^{2}-0.003 M_{1}^{2}+0.091 m_{0}^{2}+0.099 A_{t_0}^{2}
\nonumber \\
&&-0.345 M_{3}A_{t_{0}}-0.077 M_{2}A_{t_{0}}-0.012 M_{1}A_{t_{0}}\nonumber \\
&&+0.22 M_{3}M_{2}+0.031 M_{3}M_{1}+0.006M_{2}M_{1}, \label{h1}
\end{eqnarray}
 \begin{eqnarray}
 A_{t}\left( M_{Z}\right) & \simeq &-2.012 M_{3}-0.252 M_{2}-0.0316 M_{1}+0.273 A_{t_0},
 \label{h2}
 \end{eqnarray}
\begin{eqnarray}
m_{Q_{t}}^{2}\left( M_{Z}\right)
& \simeq &5.41 M_{3}^{2}+0.392 M_{2}^{2}-0.007 M_{1}^{2}+0.64 m_{0}^{2}-0.033 A_{t_0}^{2}
\nonumber \\
&&+0.115 M_{3}A_{t_{0}}+0.026 M_{2}A_{t_{0}}+0.004 M_{1}A_{t_{0}}\nonumber \\
&&-0.072 M_{3}M_{2}-0.01 M_{3}M_{1}-0.002 M_{2}M_{1}, \label{h3}
\end{eqnarray}
\begin{eqnarray}
m_{U_{t}}^{2}\left( M_{Z}\right)
& \simeq &4.52 M_{3}^2-0.188 M_{2}^2+0.044 M_{1}^2+0.273 m_{0}^2-0.066 A_{t_0}^2 \nonumber\\
&&+0.23 M_{3}A_{t_0}+0.051 M_{2}A_{t_0}+0.008 M_{1}A_{t_0} \nonumber\\
&&-0.145 M_{3}M_{2}-0.0208 M_{3}M_{1}-0.004 M_{2} M_{1}.
 \label{h4}
\end{eqnarray}

\begin{figure}[th]
\centering  \includegraphics[angle=0, width=17cm]{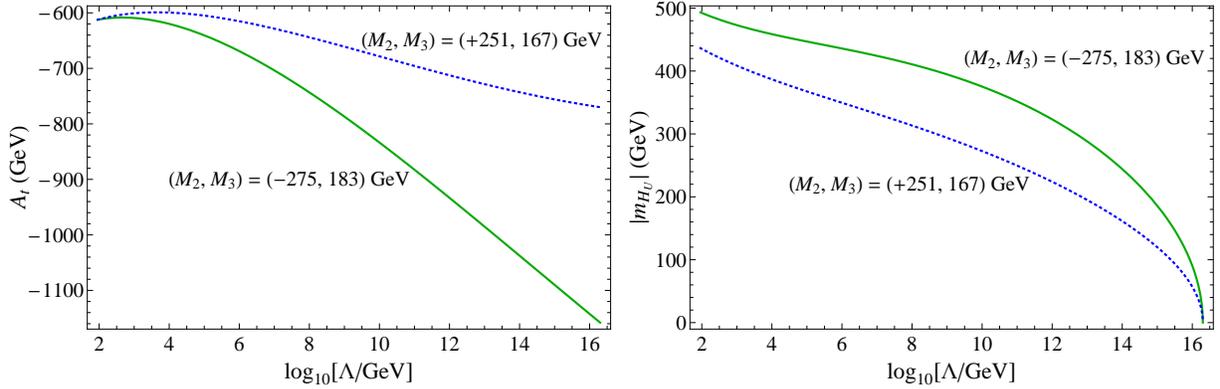}
\caption{Evolution of $A_t$ and $\vert m_{H_{u}} \vert$ for
$\vert M_2/M_3 \vert =1.5$ at $M_G$, $M_S=250$ GeV  and
$A_t/M_S=-\sqrt6$ at EW scale, which corresponds to
$m_h=119$ GeV. } \label{evol1}
\end{figure}

We observe from Eq.(\ref{h1}) that in order to  reduce the absolute
value of $m_{H_{u}}^{2}\left( M_{Z}\right)$ ( which, for our purpose,
can be regarded as a measure of fine tuning), it is useful to have comparable values
for the first two terms. This suggests the need for non-universal
gaugino masses at $M_G$, with $\vert M_2 \vert > \vert M_3 \vert$.
It turns out that the relative sign between $M_3$ and $M_2$
plays an important role. Since $M_1$ appears with relatively small
coefficients in Eq.(\ref{h1})--(\ref{h4}), we can neglect terms
containing it in the qualitative discussion. However, all terms are
included in the numerical analysis.

From Eq.(\ref{h2}) it follows that opposite signs for $M_2$ and
$M_3$ help reduce the absolute value of $A_t(M_Z)$. However, this
reduces  $m_h$, and to compensate for this one should increase the
value of $M_3$ at $M_G$. This, in turn, increases the absolute value
of $m_{H_{u}}^{2}\left( M_{Z}\right)$. In order to show how
$m_{H_{u}}^{2} \left( M_{Z}\right)$ depends on the sign accompanying
the ratio $M_3/M_2$ we present in Figure \ref{evol1}, the RGE running
of $A_t$ and $m_{H_{u}}^2$ by taking $\vert M_2/M_3 \vert =1.5$ (as an example),
with $m_h=119$ GeV, $M_S=250$ GeV and $A_t(M_Z)/M_S = - \sqrt6$. 
In order to generate the same $A_t$ value at low scale,
for $M_S = 250$ GeV we obtain $M_3=167.4$ GeV and $A_{t_0} = - 825$ GeV with
$M_2/M_3 = +1.5$, while $M_3=183.3$ GeV and $A_{t_0} = - 1440$ GeV for 
$M_2/M_3 = -1.5$. Next, with the same asymptotic value for $m_{H_u}^2$ and using the above results 
in its RGE running (right panel in Figure \ref{evol1}), smaller values 
of $m_{H_{u}}^2$ are realized for $M_2/M_3=+1.5$. Thus,
in order to reduce the amount of fine tuning,
the following two conditions should be met, namely
$\vert M_2/M_3 \vert > 1$ and the ratio $M_2/M_3$ has to be positive.

\begin{figure}[h]
\centering
\includegraphics[angle=0,width=8.5cm]{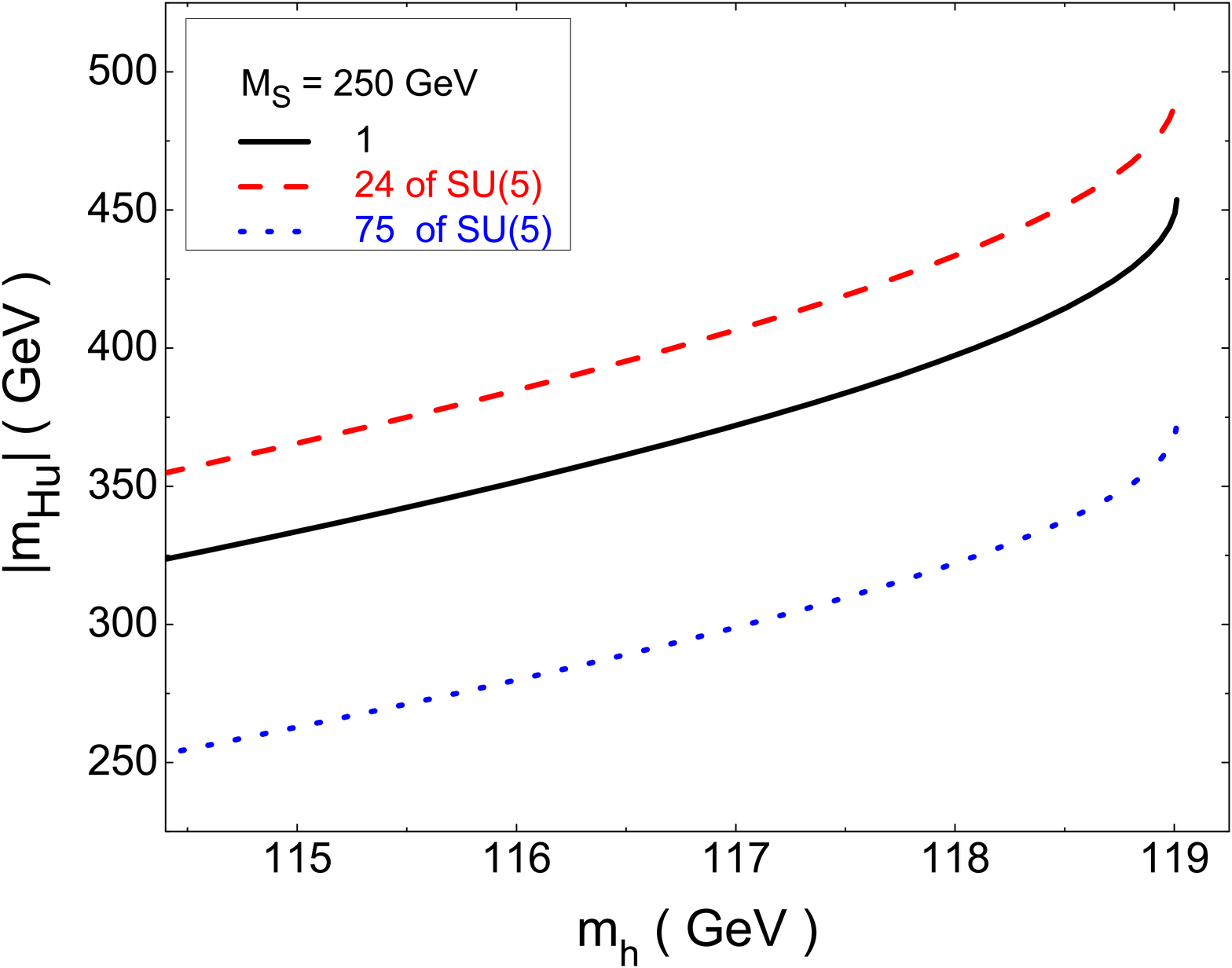} \hspace{-1.5cm}
\includegraphics[angle=0,width=8.5cm]{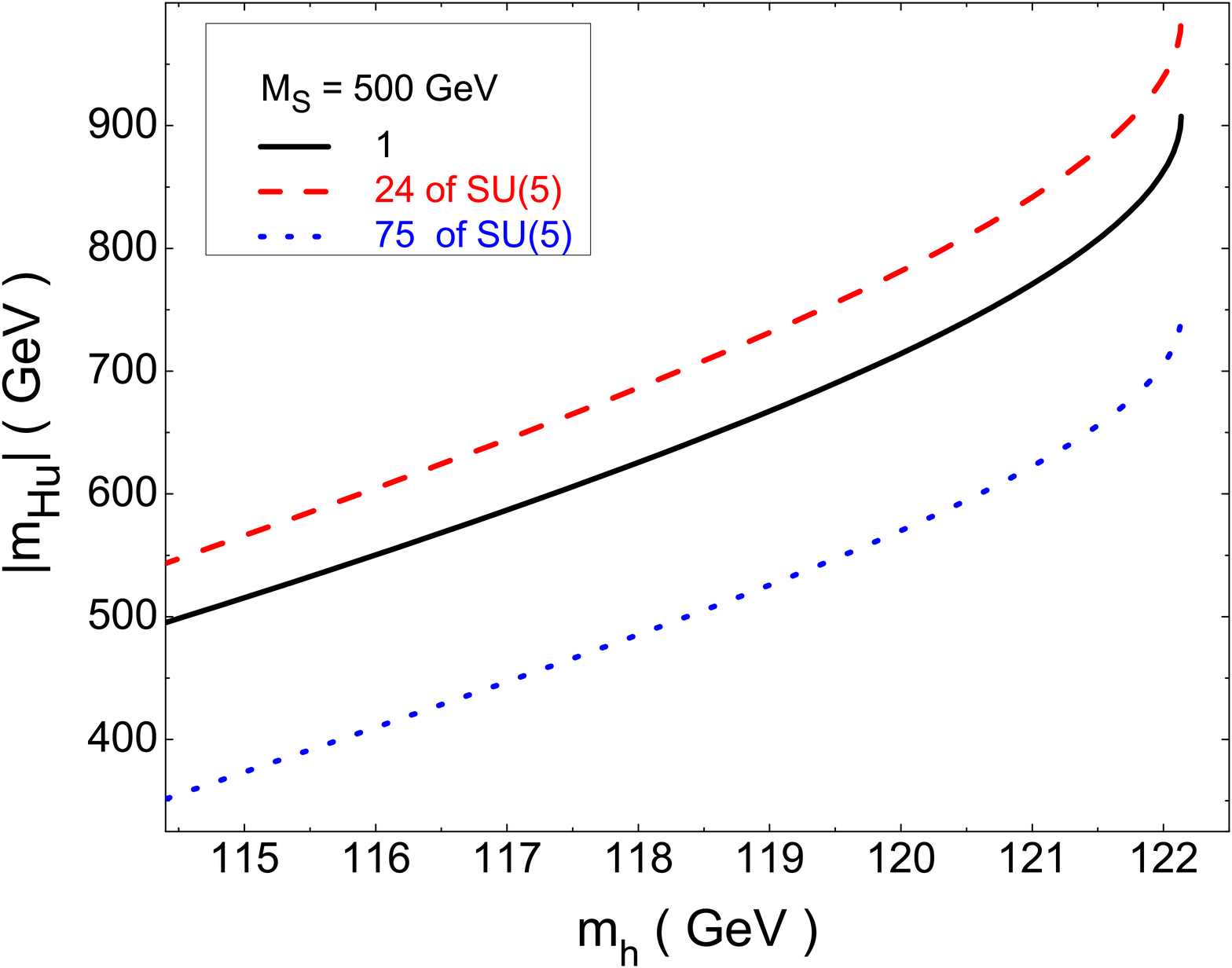} \vspace{-1.0cm}\\
\includegraphics[angle=0,width=8.5cm]{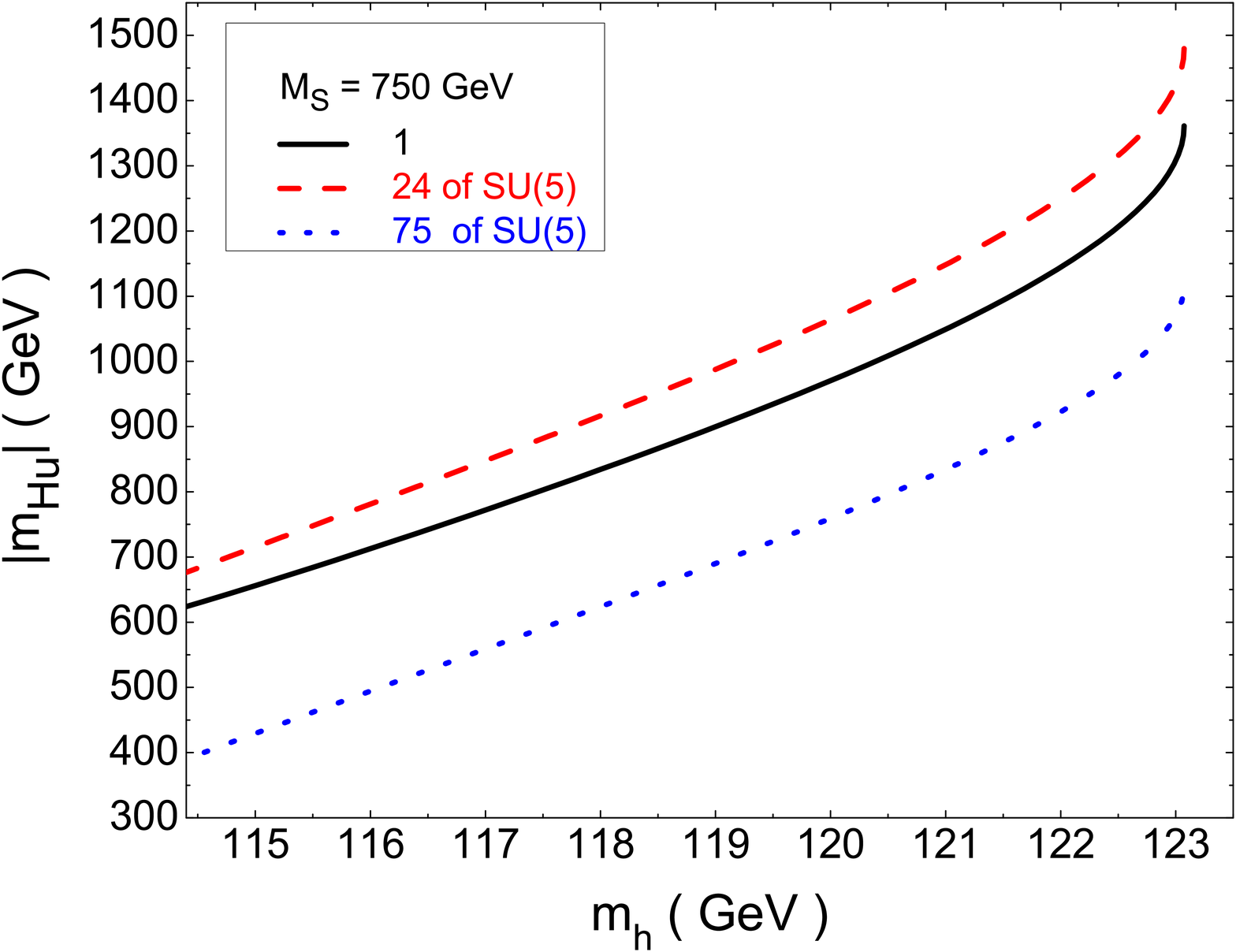}\hspace{-1.5cm}
\includegraphics[angle=0,width=8.5cm]{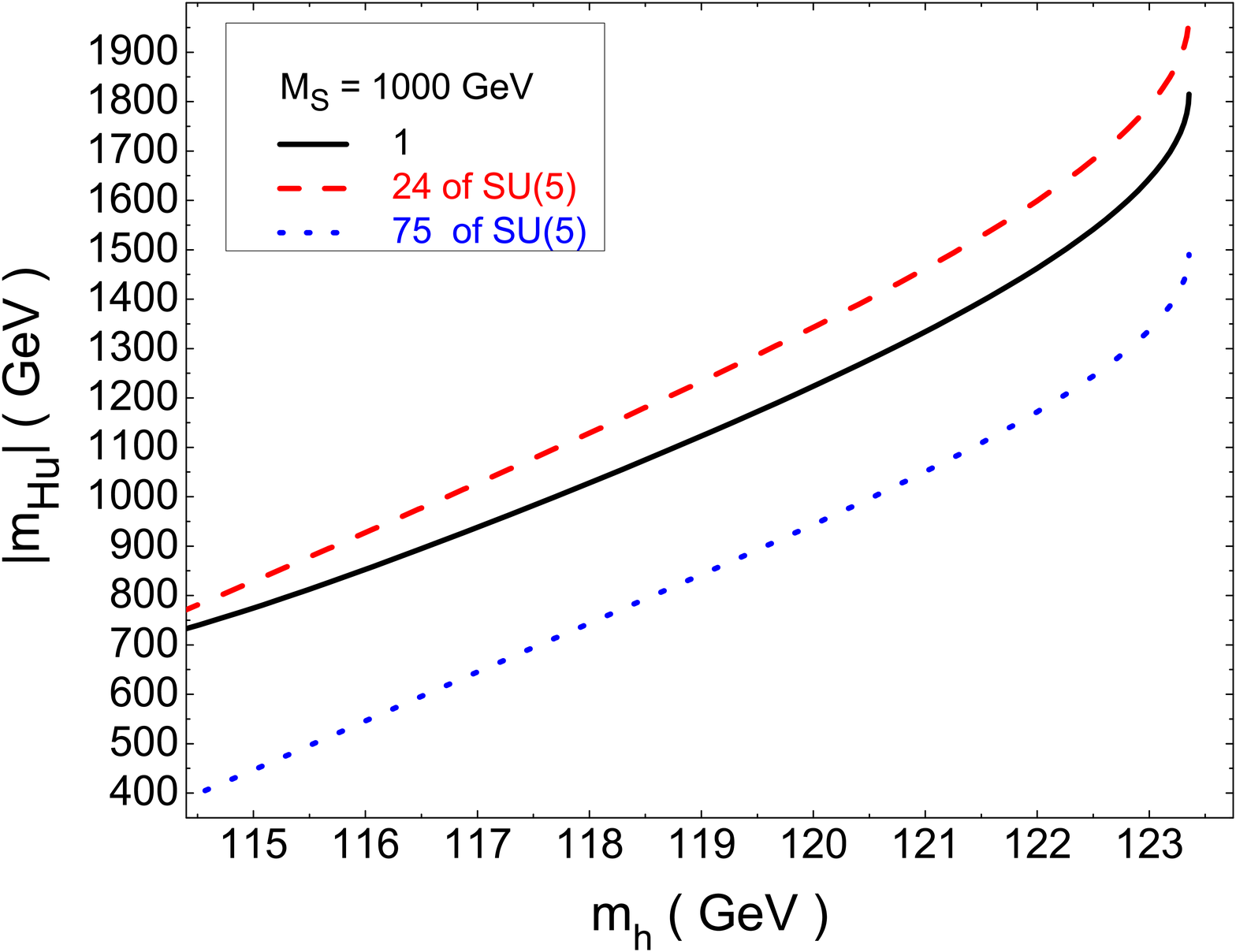}
\caption{$\vert m_{H_{u}} \vert$ vs $m_h$ for different gaugino mass
ratios presented in Table 1, with $M_S$ varying from 250 GeV to 1
TeV.  $SO(10)$ breaking with 54-plet via $G_{422}$ channel
yields the same results as the $SU(5)$ breaking with 
24-plet}. \label{MHU}
\end{figure}
This observation helps us to explain the results presented in Figure
\ref{MHU} for various values of the gaugino mass ratios and $M_S$
given in Table 1. With non-universal gaugino masses the little
hierarchy problem is somewhat less severe compared to the CMSSM  if
the 75  dimensional representation of $SU(5)$ is employed. This
case has $M_2>M_3$ at $M_G$, with the ratio $M_2/M_3$ positive. Note
that the 75-plet  of $SU(5)$ can also lead to a natural solution of
the doublet-triplet splitting problem \cite{mp}. Note, however, that the little
hierarchy problem becomes more severe compared to CMSSM if we employ 
the 24-plet of $SU(5)$ or 54-plet of $SO(10)$.

We will compare the fine tuning measures in the CMSSM with the $SU(5)$
model containing the $75$ representation. Employing the
semi-analytic result for $M_{Z}^2$ and Eq.(\ref{ftdef}) with $(M_1,M_2,M_3)
=(-5,3,1)\,M$, we can express the $\Delta_X$ parameters for the $SU(5)$ model as follows:
\begin{eqnarray}
\Delta_{M}
&=& 2.32 \,\hat{M}^2 - 0.51 \,\hat{M} \hat{A}_{t_0}, \label{eq:deltaM} \\
\Delta_{A_{t_0}}&=& 0.2 \,\hat{A}_{t_0}^2 - 0.51 \, \hat{M} \hat{A}_{t_0},  \\
\Delta_{\mu_0} &=& -2.04\,\hat{\mu}_0^2,
\end{eqnarray}
where
\begin{eqnarray}
\hat{M_Z}^2 &=& 2.32\, \hat{M}^2 + 0.2\, \hat{A}_{t_0}^2 - 1.02\, \hat{M} \hat{A}_{t_0} -2.04\, \hat{\mu}_0^2,
\label{eq:mz1}
\end{eqnarray}
and $\hat{M}_a = M_a/M_Z$, $\hat{A}_{t_0} = A_{t_0}/M_Z$ and $\hat{\mu}_0 = \mu_0 / M_Z$.

\begin{table}
\begin{center}
\begin{tabular}{|c||c|c|c|c||c|c|c|c|}
\hline
&\multicolumn{4}{c||}{CMSSM}& \multicolumn{4}{c|} {{$SU(5)$ (with \bf 75}-plet)} \\
\hline
$r$  & 1 & 1 & 1 & 1 & 3 & 3 & 3 & 3  \\
\hline
$M_{S}$  & 250 & 500 & 750 & 1000 & 250 & 500 & 750 & 1000  \\
\hline
$\left( \frac{100}{max(\vert\Delta_X \vert )} \right)\% $ & $4\%$ & $1.5\%$ & $0.8\%$ & $0.5\%$ & $5.4\%$ & $3\%$ & $1.6$\% & $1\%$ \\
\hline
$-\frac{A_t(M_Z)}{M_{S}}$ & $2.4$ & $1.8$ & $1.7$ & $1.6$ &  $2.4$ & $1.8$ & $1.7$ & $1.6$ \\
\hline
$M_3$ &  167 & 279 & 398 & 518 & 159 & 262 & 372 & 477 \\
\hline
$M_2$ &  167 & 279 & 398 & 518 & 477 & 785 & 1117 & 1432  \\
\hline
$M_1$ &  167 & 279 & 398 & 518 & -795 & -1309 & -1862 & -2387  \\
\hline
$\mu(M_Z)$ &  443 & 664 & 898 & 1121 & 362 & 522 & 687 & 840  \\
\hline
$m_{Q_t}(M_Z)$ &  342 & 605 & 879 & 1153 & 426 & 724 & 1040 & 1354  \\
\hline
$m_{U_t}(M_Z)$ &  183 & 413 & 640 & 867 & 147 & 345 & 541 & 739 \\
\hline
\end{tabular}
\end{center}
\caption{Fine tuning in CMSSM  with universal gaugino masses and in $SU(5)$ with non-universal gaugino masses. We set $m_h = 119$ GeV and tan$\beta$ = 10. All masses are in GeV.}\label{tab:su5}
\end{table}

Using Eqs.(\ref{eq:deltaM})-(\ref{eq:mz1}) and the
semi-analytic relations in Eqs. (\ref{h0})-(\ref{h4}), we calculate the
degree of fine tuning and the SUSY parameters $M_{1,2,3}$, 
$m_{Q_t} (M_Z)$, $m_{U_t} (M_Z)$, $\mu (M_Z)$ and $A_t (M_Z)$  in the CMSSM and
$SU(5)$ with 75-plet, for different values of $M_S$. We choose $m_h = 119$ GeV as an example. The 
results presented in Table \ref{tab:su5} show a marginal
improvement in fine tuning in $SU(5)$ relative to the CMSSM case.

\section{Little Hierarchy Problem in $G_{422}$}

In this section we will show that the little hierarchy problem is
largely resolved if the MSSM is embedded in $G_{422}$ \cite{Pati:1974yy}. It
seems natural to assume that in $G_{422}$ the asymptotic gaugino masses
associated with $SU(4)_c, SU(2)_L$, and  $SU(2)_R$ are three
independent parameters. This number can be reduced from three to two
in the presence of C (or D) parity \cite{cparity}. (C parity
interchanges left and right and simultaneously conjugates the
representations). For instance, C invariance requires that the
$SU(2)_L$ and $SU(2)_R$ gauge couplings are equal at the $SU(2)_R$
breaking scale, which we identify with the GUT scale $M_G$. Applying
C-parity to the gaugino sector we can realize  $SU(2)_L$ and
$SU(2)_R$ gaugino mass unification at $M_G$, but the $SU(3)_c$
asymptotic gaugino mass is still independent. $G_{422}$ symmetry and
C-parity imply the following asymptotic relation among the MSSM
gaugino masses:
\begin{equation}
M_1 = {\frac{2}{5}}M_3+{\frac{3}{5}}M_2, \label{PSMR}
\end{equation}
where $M_3, M_2, M_1$ denote the $SU(3)_c, SU(2)_L$ and $U(1)_Y$
asymptotic gaugino masses respectively. In deriving Eq.(\ref{PSMR}), we used the relation $Y=\sqrt{\frac{2}{5}} (B-L)+\sqrt{\frac{3}{5}} I_{3R}$,
where $B-L$ and $I_{3R}$ are the diagonal generators of $SU(4)_c$ and
$SU(2)_R$ respectively.

\begin{figure}[h]
\centering  \includegraphics[angle=0, width=18cm]{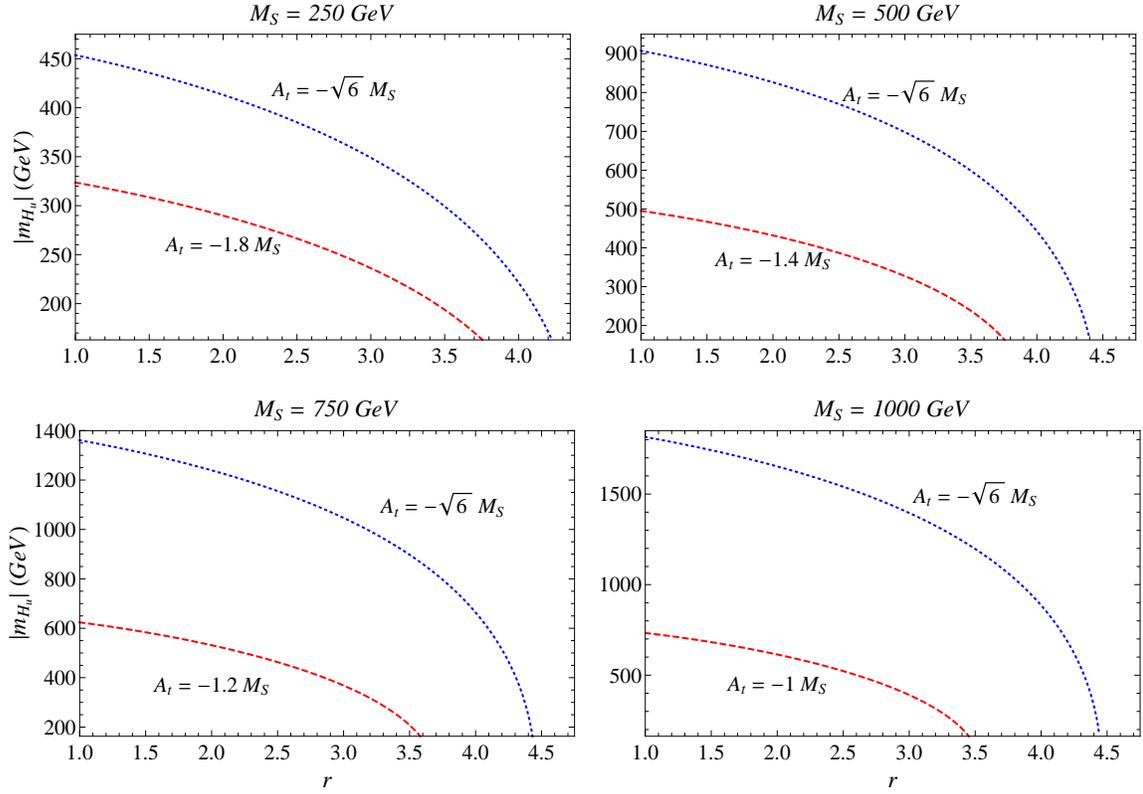}
\caption{$\vert m_{H_u} \vert$ vs $r$ for $M_S =$ 250 GeV, 500 GeV,
750 GeV and 1000 GeV, with $M_1 =
{\frac{2}{5}}M_3+{\frac{3}{5}}M_2$ and $M_2 = r M_3$. In all four
plots the lower (dashed red) curve corresponds to $m_h$ = 114.4 GeV,
while the upper (dotted blue) curve corresponds to the maximum Higgs mass
$m_h = 119$ GeV, $122$ GeV, $123$ GeV and $123$ GeV respectively.} \label{MHUr}
\end{figure}

Following our earlier discussion about the fine tuning measure (defined in Eq. (\ref{ftdef})),
the three relevant $\Delta_X$ parameters are given by
\begin{eqnarray}
\Delta_{M}
&=& 0.4 \,(4.3-r)\,(3.1+r)\, \hat{M}^2 - 0.5\,(0.7+0.17\,r)\, \hat{M} \hat{A}_{t_0}, \\
\Delta_{A_{t_0}}&=& 0.2\,\hat{A}_{t_0}^2 - 0.5\,(0.7+0.17\,r)\, \hat{M} \hat{A}_{t_0},  \\
\Delta_{\mu_0} &=& -2.04\,\hat\mu_0^2,
\label{eq:deltas2}
\end{eqnarray}
with
\begin{eqnarray}
\hat{M_Z}^2 &=& 0.4 \,(4.3-r)\,(3.1+r)\, \hat{M}^2 + 0.2 \,\hat{A}_{t_0}^2 - (0.7+0.17\,r)\, \hat{M} \hat{A}_{t_0} -2.04\, \hat{\mu}_0^2,
\label{eq:mz2}
\end{eqnarray}
where $r \equiv M_2/M_3$ and $M \equiv M_3$. Thus, $(M_1,M_2,M_3)=(\frac{2+3\,r}{5},r,1)\,M$.
We expect to reduce  $\vert m_{H_{u}}\vert$ for larger values of
$r$. This can be seen from a comparison between $G_{422}$ model and
CMSSM shown in Table \ref{tab:gps2}. For $M_S \sim$ TeV the problem
is largely overcome with $10\%$ fine tuning in $G_{422}$  versus a fine
tuning of $0.5\%$ in the CMSSM. This is further exemplified in
Figure \ref{MHUr}. With  $r$ in the interval $3.5<r<4.5$, and $M_S=250$ GeV (top-left panel in Figure \ref{MHUr}), the EW symmetry breaking condition requires only $10\%$ cancellation,
which may be regarded as modest amount of fine tuning.

\begin{table}[]
\begin{center}
\begin{tabular}{|c||c|c|c|c||c|c|c|c|}
\hline
&\multicolumn{4}{c||}{CMSSM} & \multicolumn{4}{c|}{$G_{422}$} \\
\hline
$r$  & 1 & 1 & 1 & 1 & 3.71 & 4.05 & 4.104 & 4.101 \\
\hline
$M_{S}$  & 250 & 500 & 750 & 1000 & 250 & 500 & 750 & 1000 \\
\hline
$\left( \frac{100}{\vert\Delta_M \vert} \right)\% $ & $4\%$ & $1.5\%$ & $0.8\%$ & $0.5\%$ & $10\%$ & $10\%$ & $10$\% & $10\%$ \\
\hline
$-\frac{A_t(M_Z)}{M_{S}}$ & $2.4$ & $1.8$ & $1.7$ & $1.6$ &  $2.4$ & $1.8$ & $1.7$ & $1.6$  \\
\hline
$M_3$ &  167 & 279 & 398 & 518 & 170 & 284 & 404 & 520  \\
\hline
$M_2$ &  167 & 279 & 398 & 518  & 630 & 1151 & 1656 & 2131 \\
\hline
$M_1$ &  167 & 279 & 398 & 518 & 446 & 805 & 1155 & 1486  \\
\hline
$\mu(M_Z)$ &  443 & 664 & 898 & 1121  & 259 & 230 & 186 & 110 \\
\hline
$m_{Q_t}(M_Z)$ &  342 & 605 & 879 & 1153 & 534 & 956 & 1375 & 1778  \\
\hline
$m_{U_t}(M_Z)$ &  183 & 413 & 640 & 867 & 117 & 262 & 409 & 563 \\
\hline
\end{tabular}
\end{center}
\caption{Fine tuning measure in CMSSM  with universal gaugino masses and in $G_{422}$ with non-universal gaugino masses. We set $m_h = 119$ GeV and tan$\beta$ = 10. All masses are in GeV.}\label{tab:gps2}
\end{table}

Before concluding, we present two examples of the Higgs and
sparticle mass spectra which is predicted from the $G_{422}$ model with
tan$\beta \sim 10 $. The data presented in Table \ref{table4} is generated using the
software program SuSpect \cite{Djouadi:2002ze}, and for these two
examples the MSSM parameter $\mu$ is close to 200 GeV, only a factor
of two larger than $M_Z$. The data is consistent with
the low energy constraints such as $m_h \geq 114.4$
GeV, lightest chargino mass $m_{\tilde{\chi}^{\pm}} > 103.5$ GeV, and
$2.85 \times 10^{-4} \leq BR(B \rightarrow X_{s} \gamma)\leq 4.24
\times 10^{-4} \, (2\sigma)$. 
The lightest neutralino (LSP) mass in this table has the right magnitude
to account for the recent results reported by the PAMELA experiment
\cite{Gogoladze:2009kv}, provided one assumes that the LSP is not absolutely
stable but decays primarily into leptons with a lifetime $\sim 10^{26}$ sec \cite{Adriani:2008zr}.
Finally, it was recently shown in \cite{Gogoladze:2009ug} 
that third family Yukawa unification and neutralino dark matter are fully consistent in a 
framework with $G_{422}$ compatible non-universal gaugino masses.

\begin{table*}[t]
\centering
\begin{tabular}{lcc}
\hline \hline
          & Point 1 & Point 2    \\
\hline
$M_{1}$ & 891   & 1486     \\
$M_{2}$ & 1241   & 2071   \\
$M_{3}$ & 365   & 609    \\
$m_{0} $ &  100  & 100    \\
$\tan\beta$ & 10   & 10  \\
$A_0$ &  -260  & -485  \\
$\mu$   & 190   &  190  \\
\hline
$m_h$          & 115.5   & 119.0      \\
$m_H$          & 834.2   & 1351.1    \\
$m_A$          & 834.0   & 1350.9    \\
$m_{H^{\pm}}$  & 838.1   & 1353.6    \\
\hline
$m_{\tilde{\chi}^{\pm}_{1,2}}$
& 189.9, 999.6 & 193.2, 1673   \\
$m_{\tilde{\chi}^0_{1,2,3,4}}$
& 181.9, 195.8, 386.1, 999.6
& 189.4, 196.7, 649.3, 1673  \\
$m_{\tilde{g}}$ & 866.2 & 1390    \\
\hline $m_{{\tilde{u}}_{L,R}}$
& 1081, 772.4 & 1752, 1228  \\
$m_{\tilde{t}_{1,2}}$
& 339.1, 995.6 & 510.4, 1588  \\
\hline $m_{{\tilde{d}}_{L,R}}$
& 1083, 571.1 & 1753, 1190  \\
$m_{\tilde{b}_{1,2}}$
& 739.9, 973.2 & 1172, 1576   \\
\hline
$m_{\tilde{\nu}_{1,2,3}}$
& 823.3, 823.3, 820.8 & 1355, 1355, 1351    \\
\hline
$m_{{\tilde{e}}_{L,R}}$
& 826.9, 346.3  & 1357, 557.1   \\
$m_{\tilde{\tau}_{1,2}}$
& 333.8, 824.5  & 536.4, 1353  \\
\hline \hline
\end{tabular}
\caption{ Sparticle and Higgs masses (in GeV), with $M_t=172.6$ GeV, $r = 3.4$
and $10\%$ fine tuning.} \label{table4}
\end{table*}

\newpage

\section{\protect\bigskip Conclusion}
We have argued that the little hierarchy problem is ameliorated in supersymmetric 
models based on the gauge symmetry $G_{422} \equiv SU(4)_c \times SU(2)_L \times SU(2)_R$
supplemented by a discrete left-right symmetry (C-parity). We have also investigated
$SU(5)$ and $SO(10)$ models in which non-universal gaugino
masses can arise from dimension five operators. Based on these considerations some 
benchmark points depicting the Higgs and sparticle masses in $G_{422}$ are highlighted.

\newpage

\section*{Acknowledgments}
We thank K.S. Babu and Rizwan Khalid for valuable discussions. This work is
supported in part by the DOE under grant \# DE-FG02-91ER40626 (Q.S.
I.G. and M.R.), GNSF grant 07\_462\_4-270 (I.G.) and the Bartol
Research Institute (M.R.).

\end{document}